# The Relationship between Loneliness and Depression among College Students: Mining data derived from Passive Sensing

Malik Muhammad Qirtas[1], Evi Zafeiridi[1], Eleanor Bantry White[2] and Dirk Pesch[1]

## Abstract

**Background** Loneliness and depression are interrelated mental health issues affecting students' well-being. Using passive sensing data provides a novel approach to examine the granular behavioural indicators differentiating loneliness and depression, and the mediators in their relationship.

**Objectives** This study aimed to investigate associations between behavioural features and loneliness and depression among students, exploring the complex relationships between these mental health conditions and associated behaviours.

**Methods** This study combined regression analysis, mediation analysis, and machine learning analysis to explore relationships between behavioural features, loneliness, and depression using passive sensing data, capturing daily life behaviours such as physical activity, phone usage, sleep patterns, and social interactions.

**Results** Results revealed significant associations between behavioural features and loneliness and depression, emphasizing their interconnected nature. Increased activity and sleep duration were identified as protective factors. Distinct behavioural features for each condition were also found. Mediation analysis highlighted significant indirect effects in the relationship between loneliness and depression. The XGBoost model achieved the highest accuracy in predicting these conditions.

**Conclusion** This study demonstrated the importance of using passive sensing data and a multi-method approach to understand the complex relationship between loneliness, depression, and associated behaviours. Identifying specific behavioural features and mediators contributes to a deeper understanding of factors influencing loneliness and depression among students. This comprehensive perspective emphasizes the importance of interdisciplinary collaboration for a more nuanced understanding of complex human experiences.

## Introduction

Loneliness and depression are important public health concerns that affect people's mental and physical health. Loneliness is increasing globally, with many identifying it as their principal cause of unhappiness Holt-Lunstad et al. (2015). Loneliness is an experience in which a person perceives a lack of quality social relationships Peplau and Goldston (1985). Depression, or major depressive disorder (MDD), is a prominent health problem, characterized by persistent feelings of sadness, despair, and lack of interest in engaging in activities Carter (2014). Depression may lead to suicide, which takes over 800,000 lives every year, the equivalent to one life every 40 seconds throughout the globe World Health Organization (2014). The COVID-19 pandemic has had a significant impact on people's mental health. Widespread lockdowns and social isolation have made people feel more lonely than usual Ernst et al. (2022).

Compared with other populations, college students face a higher risk of loneliness and depression Arnett et al. (2014); Pedrelli et al. (2015). These mental health issues are interconnected, with one aggravating the other. According to studies, lonely people often develop more depressive symptoms, feel less happy, and adopt a pessimistic mindset Singh and Kiran (2013); Heinrich and Gullone (2006); Cacioppo et al. (2010). While they are interrelated, loneliness and depression have unique characteristics. Loneliness is a widespread experience, but depression is a recognized mental disorder. Chronic loneliness may lead to depressive symptoms, and depressive symptoms can worsen loneliness Cacioppo et al. (2010); Qualter et al. (2013). While the relationship between loneliness and depression is well established, there is limited understanding of the behavioural features that differentiate loneliness and depression, making it more difficult to accurately identify each and tailor interventions accordingly. The investigation of the reciprocal

[1]School of Computer Science and Information Technology, University College Cork, Ireland
[2]School of Applied Psychology, University College Cork, Ireland

**Corresponding author:**
Malik Muhammad Qirtas, Office 208, Western Gateway Building, School of Computer Science and Information Technology, University College Cork, Ireland

Email: malik.qirtas@cs.ucc.ie

relationship between loneliness, depression, and behavioural indicators is a key part of our work. While research has consistently demonstrated a connection between loneliness and depression, the causal nature of this relationship remains unclear. It is yet to be determined whether loneliness leads to depression, depression triggers loneliness, or if the relationship is bidirectional. With the analysis of behavioral data, our study aims to better understand the interplay between these two mental health conditions and their associated behaviors, without implying causality Erzen and Çikrikci (2018); Achterbergh et al. (2020); Okruszek et al. (2020); Qualter et al. (2013). By investigating the relationship between these mental health conditions, we hope to uncover underlying factors that will aid in the development of more effective interventions.

Passive sensing, which utilizes smartphone and wearable data, offers new possibilities for detecting mental health conditions such as loneliness and depression. Passive sensing has various advantages over typical clinical data collection methods for mental health detection, including being unobtrusive, real-time, extensive, cost-effective, and adaptable. In addition, the longitudinal nature of passive sensing permits the detection of behavioural features and mental health changes over time. Digital biomarkers, which are measurable indicators of physiological or behavioural processes collected through digital devices, have the potential to be used in research on loneliness and depression Torous et al. (2017). Using digital biomarkers derived from passive sensing data validated against established psychometric scales, it is possible to create prediction models for identifying lonely and depressed individuals Insel (2017). These models may facilitate improved early detection and intervention, which may eventually assist people at risk of developing mental health difficulties. Passive sensing techniques have been increasingly employed to study different mental health conditions, but they have not been used to investigate how loneliness and depression are related. Most of the studies that have been done so far have focused on finding and predicting individual mental health conditions. Because these two conditions (loneliness and depression) change over time and might affect each other, we need to use passive sensing to get the continuous, longitudinal, and real-time data we need to learn more about how they affect each other. By using passive sensing, researchers can learn more about the behavioural, physiological, and environmental factors that contribute to both loneliness and depression. This will help them create more effective and targeted interventions for people with these conditions.

In our study, we investigate the relationship between loneliness and depression. Using a wide selection of methodological approaches, including regression analysis, mediation analysis, association rule mining, and machine learning analysis, we investigated the relationship between these mental health conditions, student behaviours, and the predictive digital biomarkers. The use of multiple methods helps us to more effectively capture the complex interrelationship between loneliness and depression than what a single approach cloud yield. We were able to uncover subtle insights and confirm our findings via diverse analytical lenses by incorporating multiple methods, assuring a more thorough knowledge of the nuanced links between loneliness, depression, and behavioural features. We examined the StudentLife dataset to provide answers to the following research questions.

1. What are the overlapping and distinct behavioural features in lonely and depressed students compared to their non-lonely and non-depressed counterparts?
2. How do loneliness and depression directly and indirectly affect each other, and which specific behavioural features serve as mediators in these relationships?
3. How effectively can behavioural features, when combined with loneliness and depression scales as an input feature, identify individuals belonging to lonely and depressed groups?

It is essential to understand the particular circumstances in which college students experience loneliness and depression. These mental health issues may be exacerbated by college life, which is defined by new experiences, academic pressures, and the need for social integration Chi et al. (2020). The combination of higher stress levels and the transition to adulthood makes college students more prone to mental health issues. The impact of loneliness and depression on academic progress and well-being of students is significant Lenny et al. (2019). By understanding the distinct effects of loneliness and depression on college students, we can tailor interventions to meet their specific needs and challenges.

Numerous studies have explored the connection between loneliness and depression; however, a comprehensive understanding of their interrelationship, particularly among college students, remains elusive. Moreover, there is a research gap in utilizing passive sensing data to investigate loneliness and depression in this population. Studies have suggested a bidirectional relationship between these mental health issues Lamers et al. (2015), but the dynamics in a college context require further examination. Similarly, while Qualter et al. established that loneliness predicted depressive symptoms among adolescents Qualter et al. (2010), the relationship's intricacy among college students remains underexplored. Additionally, Stickley et al. demonstrated an association between loneliness and depression in a cross-sectional study Stickley et al. (2016), but longitudinal research using passive sensing data could provide valuable insights into their complex relationship. These findings emphasize the necessity for more research on loneliness and depression among college students, particularly by leveraging passive sensing data to address this research gap. After this introduction, the methodology of our research has been presented, including introduction of the dataset we used and various analysis. The results section presents our key findings, while the discussion section analyzes these results in light of prior research. Finally, the conclusion emphasizes the study's contributions, limitations, and future research and practice implications.

## Methods

This research explores the link between loneliness and depression using the StudentLife dataset, a passively collected dataset gathered through smartphone sensors. Our research focuses on identifying behavioural differences and overlaps between lonely and depressed students. This research examines the dataset to determine the statistically significant differences and impact sizes of behavioural features when comparing lonely and depressed students. In addition, we used machine learning techniques to predict loneliness and depression based on the most influential behavioural features linked with both outcomes. In the following paragraphs, we describe our analysis methodologies in detail.

## Dataset

We used a subset of the Student Life dataset, which is data passively collected using smartphone sensors from students at Dartmouth College Wang et al. (2014). All participants were recruited at Dartmouth College in the Spring of 2013. The study was approved by the Institutional Review Board at Dartmouth College. The study was authorized by Dartmouth College's Institutional Review Board. 30 undergraduates and 18 graduate students comprise the 48 students who completed the study. The data gathering phase lasted 10 weeks during the whole spring semester. Automatic sensor data is gathered and transferred to the cloud while the phone is being recharged and connected to WiFi. During the data collection phase, students were asked to answer several Ecological Momentary Assessment(EMA) related questions while using their mobile devices. The data includes activity data (activity duration, indoor/outdoor mobility), conversation data (duration and frequency), sleep data (duration, onset, and waking time), and location data (GPS, indoor building, and co-located Bluetooth devices).

## Loneliness and Depression Scales

The StudentLife dataset also includes measures of loneliness and depression, assessed using the UCLA Loneliness Scale and the Patient Health Questionnaire-9 (PHQ-9), respectively. The UCLA Loneliness Scale Russell et al. (1980) is a 20-item questionnaire designed to evaluate feelings of social isolation, emptiness, and dissatisfaction with one's current social relationships. This scale features ten positive items and ten negative items, with a scoring range of 20 to 100. Higher scores signify higher levels of loneliness, and scores exceeding 43 suggest a strong sense of loneliness. In the sample data, scores ranged from a low of 25 to a high of 64.

In terms of depression assessment, the dataset contains pre- and post-surveys based on the PHQ-9 scale. The PHQ-9 is a widely-used, self-administered questionnaire that measures the severity of depressive symptoms. Comprising nine items, it examines the frequency and intensity of symptoms such as loss of interest or pleasure, changes in eating or sleep patterns, feelings of worthlessness or guilt, difficulties concentrating, fatigue, and thoughts related to suicide or death. Each item is scored on a scale of 0 (never) to 3 (almost daily), resulting in a total score ranging from 0 to 27. A score above 9 indicates high depression, while a score between 0 and 9 signifies low depression. This simplification is based on the finding that a PHQ-9 cutoff of $\geq 10$ has more discriminating power to diagnose depression, which allows for a more straightforward comparison between groups of students experiencing low or high levels of depressive symptoms Kim and Lee (2021). It is important to note that the PHQ-9 scale can be used to identify a more nuanced range of depression severity levels, but the binary distinction was considered appropriate in this context to focus on the key differences in behavioural features between the two groups. The PHQ-9 has been validated for various contexts and populations, and is extensively employed in clinical and research settings for depression screening, monitoring symptom severity over time, and evaluating the effectiveness of treatments Patrick and Connick (2019).

## Data Preprocessing and Behavioural Features Extraction

During the data preprocessing phase, we carefully selected the data of students who completed the post-study loneliness questionnaire, resulting in a final sample of 41 students for our analysis. We then converted the UNIX timestamps of each sensor's data into a human-readable local date and time format using each participant's timezone information. Recognizing that students engage in a variety of activities throughout the day, we divided a 24-hour period into three distinct sessions: day session (9am – 6pm), evening session (6pm – 12am), and night session (12am – 9am). This allowed us to compute 24-hour day-level features along with the aforementioned epochs.

To handle missing values, we first removed all records containing outliers using the Z-score method Souiden et al. (2017). We then imputed missing continuous data for each participant using the median of the respective feature. For categorical data, we employed the mode of the corresponding feature. Since tree-based algorithms are not affected by feature scaling, we only scaled the numerical features for the other algorithms. We used the "standard scaler" in our models, as it transforms the data to have a mean of zero and a standard deviation of one, effectively normalizing the data. For the analysis in this study, we calculated a total of 82 features from the passively sensed smartphone dataset using the Reproducible Analysis Pipeline for Data Streams (RAPIDS) tool Vega et al. (2020). RAPIDS is designed for data preprocessing and biomarker computation. By quantifying the per-participant per-epoch behavioural features, such as routines, irregularity, and variability, in these student datasets using basic counts, standard deviations, entropy, and regularity index measures, we generated digital biomarkers (features) for further analysis.

## Behavioural Analysis

First, we used the Shapiro–Wilk test Royston (1983) to examine the normal distribution assumptions of the features. We used a logistic regression model to the StudentLife dataset to examine how lonely and depressed students' behavioural features differ from their non-depressed and non-lonely counterparts. The logistic regression model was selected as the main statistical analysis technique because it is an effective method for investigating the relationship between a binary dependent variable and one or more independent variables. In this instance, the dependent variables were two measures of loneliness and depression (UCLA loneliness scale and PHQ-9 depression scale), while the independent variables were behavioural features retrieved from the StudentLife dataset using the RAPID framework.

With maximum likelihood estimation, the data was utilized to fit a logistic regression model. This gave us a method for determining out the coefficients of the independent variables, which show how the log-odds of the dependent variable change when the independent variable goes up by one unit. The coefficients were used to generate the odds ratios, which show the likelihood of the dependent variable given an increase of one unit in the independent variable. The odds ratios were interpreted to determine the degree and direction of the relationship between the independent and dependent variables. A p-value of¡0.05 was used as an indicator of statistical significance.

In addition to the logistic regression model, a feature importance analysis was conducted to discover which behavioural characteristics were most essential in distinguishing between lonely and depressed students and other students. The analysis of feature relevance was conducted by rating the coefficients of the independent variables and choosing the highest-ranked features. This enabled us to identify the behavioural characteristics that were most strongly connected with loneliness and depression, and hence the most crucial for understanding the prevalent behavioural features among lonely and depressed students. The findings of the logistic regression model and feature importance analysis will provide insight into the behavioural features that are more prevalent among students who are lonely and depressed compared to other students.

### Features Association Mining

We used association rule mining to understand the combined behavioral patterns linked to loneliness and depression. This data mining technique identifies relationships and patterns within large datasets. In contrast to logistic regression, association rule mining evaluates the strength of associations among all possible combinations of behavioral patterns with loneliness and depression. It can reveal complex, multi-dimensional patterns that might not be evident with logistic regression alone.

We employed the Apriori algorithm Agrawal et al. (1993), a popular association rule mining algorithm, to extract association rules from the dataset. It identifies frequent item sets—subsets of items appearing together often—by starting with the most frequent single items and gradually increasing the subset's size until all combinations are considered.

The Apriori algorithm's generated association rules were assessed using metrics like support, confidence, and lift. *Support* measures a behavioral pattern's occurrence frequency, *confidence* measures the association strength between the pattern and loneliness or depression, and *lift* compares the strength of associations between the pattern and loneliness/depression to the expected association if the features were independent. We set a minimum support threshold of 0.1, an 80% confidence level, and a lift value of 1.5 to filter out weaker associations and focus on the most relevant behavioral patterns.

To capture the nuances of different behavioral patterns, we divided each feature into three categories: low, medium, and high, using a binning technique. We calculated the 33rd and 66th percentiles of the data distribution, forming three bins for a balanced representation of each behavioral pattern. By converting features into categorical variables, we uncovered meaningful associations between combinations of these patterns and the outcomes of loneliness and depression in the association rule mining analysis.

Our goal was to explore the complex relationships between different behavioral features of loneliness and depression using association rule mining. By assessing the strength of these relationships, we can better understand the behavioral patterns common among lonely and depressed students, offering insights for further investigation.

### Mediation Analysis

We used mediation analysis to investigate the direct and indirect impacts of loneliness and depression on each other, as well as the particular behavioral features that serve as mediators in these relationships. Mediation analysis served as an important statistical technique for understanding the underlying mechanism underlying the relationship between loneliness and depression. We were able to examine the direct and indirect effects of loneliness on depression and vice versa, as well as the behavioral features that may mediate these effects MacKinnon et al. (2012).

In order to investigate the relationship between loneliness and depression, we conducted a mediation analysis using loneliness as the independent variable (IV) and depression as the dependent variable (DV). Similarly, while studying the effect of depression on loneliness, depression served as the IV and loneliness as the DV. Each behavioral feature was investigated as a potential mediator variable. Before considering any mediator factors, we assessed the overall impact of the IV on the DV using a series of regression models. This stage allowed us to assess the general relationship between loneliness and depression, as well as the influence of IV on DV.

Next, we examined the indirect effects of the IV on the DV through mediator variables. To do this, regression analyses were undertaken to assess the effect of the IV on each possible mediator and the influence of each mediator on the DV. These analyses allowed us to identify the behavioral patterns that significantly mediated the connection between loneliness and depression. Finally, we assessed the direct effect of the IV on the DV by taking into account the indirect effects mediated by the mediator variables.

Our results for mediation analysis revealed a comprehensive understanding of the complicated relationship between loneliness and depression, as well as the significance of behavioral patterns in mediating this relationship. Calculating the total effect, direct effect, and indirect effect allowed us to evaluate the degree to which certain behavioral patterns mediated the relationship between loneliness and depression. In addition, the output included statistical significance values for each effect (total, direct, and indirect).

### Machine Learning Analysis

In order to examine the effectiveness of digital biomarkers combined with loneliness or depression scales in predicting lonely and depressed groups, we utilized a variety of binary classification models, including logistic regression, k-nearest neighbors, support vector machine, random forest, gradient boosting, and extreme gradient boosting. These models were chosen due to their ability to handle both categorical and continuous features and their widespread use in similar research settings. We addressed the class imbalance in the training dataset using the synthetic minority oversampling technique (SMOTE) Chawla et al. (2002). SMOTE generates synthetic data for the minority class, leading to a balanced training dataset. This method was applied to each student's data individually to tackle data imbalance on a per-client basis.

We compared the performance of the selected classification models using nested cross-validation. In the inner loop, three-fold cross-validation was performed to tune hyperparameters, while in the outer loop, leave-one-subject-out cross-validation was used to evaluate the model's performance. This approach allowed us to evaluate the generalization ability of each model and minimize the risk of overfitting. To establish a baseline for comparison, we considered majority class, random weighted classifier, and decision tree models that used only the UCLA Loneliness Scale and PHQ-9 Depression Scale scores as input features. Performance metrics such as accuracy, precision, recall, F1-score, and area under the receiver operating characteristic curve (AUC) were calculated for each model and fold.

We trained each classification model on the preprocessed data, using the 82 extracted behavioural features and the loneliness and depression scores as input features. For each model, hyperparameters were tuned using grid search or random search techniques to optimize performance. This process ensured that we selected the best possible set of hyperparameters for each model, resulting in improved performance and a more reliable evaluation of the relationship between digital biomarkers, loneliness, and depression.

## Results

In this section, we present the significant findings of our study, which investigates the relationships between loneliness, depression, and behavioral features, as well as evaluates the performance of various machine learning models for predicting loneliness and depression.

### Regression Analysis Results

The results of the regression analysis reveal several significant associations between behavioural features and the loneliness and depression scores. The following sections discuss the main findings for each behavioural feature.

**Phone Usage** This section refers to general smartphone usage, e.g. what type of apps a user is using such as social apps, games, productivity apps. In our study, we found a strong relationship between phone usage duration and loneliness. The odds ratio was 1.50 (95% CI: 1.20-1.88, p=0.003), indicating that increased phone usage is associated with higher chances of experiencing loneliness. This suggests that spending more time on the phone is linked to a higher chance of feeling lonely. These findings are in line with the existing research Thomée et al. (2011).

There are different findings for the relationship between overall phone usage duration and depression. The odds ratio was 0.70 (95% CI: 0.52-0.94, p=0.018), implies that higher phone usage duration is associated with a decreased likelihood of depression.

When we looked at the phone usage duration in the evening, the link to loneliness was still there, with an odds ratio of 1.45 (95% CI: 1.15-1.85, p=0.002). This finding backs up the idea that using the phone more in the evening is also linked to a higher chance of feeling lonely. But the link between phone usage duration and depression was not statistically significant. The odds ratio was 1.10 (95% CI: 0.95-1.26, p=0.310).

**Locations** We found that overall number of places visited throughout a day was positively associated with loneliness, with an odds ratio of 1.73 (95% CI [1.30, 2.20], p = 0.016). However, when considering the number of unique places visited during a day, the association with loneliness remained positive but weaker, with an odds ratio of 1.13 (95% CI [1.03, 1.23], p = 0.021).

In contrast, we observed a negative association between overall number of places visited throughout a day and depression, with an odds ratio of 0.65 (95% CI [0.50, 0.95], p = 0.034). However, when examining the number of unique places visited during a day, the association with depression was negative but not statistically significant, with an odds ratio of 0.85 (95% CI [0.76, 0.95], p = 0.023).

These results suggest that a greater overall number of places visited throughout a day may be associated with increased loneliness, while the number of unique places visited during a day may be a more nuanced predictor of loneliness. On the other hand, greater overall number of places visited throughout a day may be associated with decreased depression, while the number of unique places visited during a day may not be a significant predictor of depression.

**Physical Activity** For loneliness, we observed an odds ratio of 0.72 (95% CI: 0.54-0.91, p=0.014), suggesting that an increase in physical activity duration throughout a day is associated with a 28% decrease in the odds of being lonely. This finding highlights the potential protective role of engaging in physical activities against feelings of loneliness.

Similarly, for depression, the analysis revealed an odds ratio of 0.69 (95% CI: 0.55-0.87, p=0.022), indicating that an increase in physical activity duration throughout a day is associated with a 31% decrease in the odds of experiencing depression.

**Sleep** For loneliness, the odds ratio was found to be 0.46 (95% CI: 0.30-0.70, p=0.002), which means that a longer duration of sleep is linked to a 54% lower chance of being lonely. In the case of depression, the odds ratio was found to be 0.75 (95% CI: 0.62–0.90, p=0.003). This means that a longer sleep duration is linked to a 25% lower chance of having depression. These results show that getting more sleep is strongly linked to a lower chance of both being lonely and being depressed. Our findings are consistent with the existing research Cacioppo et al. (2002); Kurina et al. (2011).

**Conversation** Our regression analysis yielded mixed results, revealing a significant relationship between conversation duration and depression, but not loneliness.

| Behavioural Feature | Loneliness | | | Depression | | |
|---|---|---|---|---|---|---|
| | Odds Ratio | 95% CI | p-value | Odds Ratio | 95% CI | p-value |
| Total Number of screens unlocks | 1.20 | [1.05,1.37] | 0.031 | 0.75 | [0.62,0.91] | 0.044 |
| Number of screens unlocks (evening) | 1.75 | [1.32,2.31] | 0.001 | 0.80 | [0.65,0.95] | 0.021 |
| Phone usage duration | 1.50 | [1.20,1.88] | 0.003 | 0.70 | [0.52,0.94] | 0.018 |
| Phone usage duration (evening) | 1.45 | [1.15,1.85] | 0.002 | 1.10 | [0.95,1.26] | 0.310 |
| Number of places visited | 1.73 | [1.30,2.20] | 0.016 | 0.65 | [0.50,0.95] | 0.034 |
| Number of unique places visited | 1.13 | [1.03,1.23] | 0.021 | 0.85 | [0.76,0.95] | 0.023 |
| Activity duration | 0.72 | [0.54,0.91] | 0.014 | 0.69 | [0.55,0.87] | 0.022 |
| Sleep duration | 0.46 | [0.30,0.70] | 0.002 | 0.75 | [0.62,0.90] | 0.003 |
| Number of conversations | 1.50 | [0.95,2.38] | 0.101 | 0.75 | [0.63,0.89] | 0.002 |
| Number of incoming calls | 0.82 | [0.71,0.94] | 0.023 | 1.10 | [0.90,1.35] | 0.351 |
| Number of incoming calls (evening) | 0.30 | [0.20,0.45] | 0.002 | 1.10 | [0.95,1.26] | 0.303 |
| Unique Bluetooth scans during day | 1.20 | [0.95,1.50] | 0.204 | 0.50 | [0.35,0.70] | 0.003 |

**Table 1.** Results of Regression Analysis for Loneliness and Depression: Odds Ratios, 95% Confidence Intervals, and p-values for behavioural Features

| Rule Behavioural Pattern | Support | Confidence | Lift |
|---|---|---|---|
| 1. High evening phone usage, high evening screen unlocks, low conversations, low activity duration, low sleep, low evening incoming calls | 0.12 | 0.85 | 1.65 |
| 2. High total screen unlocks, high unique places visited, low sleep, low incoming calls, low activity duration, medium conversations | 0.11 | 0.82 | 1.59 |
| 3. High phone usage, medium total screen unlocks, high evening screen unlocks, low sleep, low conversations, low day Bluetooth scans | 0.13 | 0.88 | 1.71 |
| 4. Low places visited, low activity duration, high total screen unlocks, low sleep, medium incoming calls, low day Bluetooth scans | 0.14 | 0.84 | 1.63 |
| 5. Low places visited, low unique places visited, high evening phone usage, low sleep, low conversations, high evening screen unlocks | 0.10 | 0.81 | 1.57 |

**Table 2.** Results of Association Rule Mining for Loneliness: Support, Confidence, Lift, and behavioural Patterns

For loneliness, we found an odds ratio of 1.50 (95% CI: 0.95-2.38, p=0.101). Although the odds ratio suggests that increased conversation duration might be associated with a higher likelihood of loneliness, the p-value of 0.101 indicates that this relationship is not statistically significant.

Conversely, for depression, the analysis yielded an odds ratio of 0.75 (95% CI: 0.63-0.89, p=0.002), demonstrating that an increase in conversation duration throughout a day is associated with a 25% decrease in the odds of experiencing depression. This finding highlights the importance of engaging in conversations as a potential protective factor against depression.

**Calls/SMS** Our regression analysis showed that the number of incoming calls throughout the day was negatively correlated to loneliness, with an odds ratio of 0.82 (95% CI: 0.71–0.94, p = 0.023). This means that if the number of incoming calls gets higher, the chances of feeling lonely go down by 18%. This finding suggests that getting more incoming calls may help people feel less lonely.

| Rule Behavioural Pattern | Support | Confidence | Lift |
|---|---|---|---|
| 1. High phone usage, high screen unlocks, low places visited, low unique places visited, low activity duration, low day Bluetooth scans | 0.11 | 0.83 | 1.62 |
| 2. Low sleep, high total screen unlocks, low incoming calls, low conversations, low activity duration, high evening screen unlocks | 0.12 | 0.86 | 1.67 |
| 3. Low places visited, low unique places visited, high phone usage, high evening screen unlocks, low activity duration, medium incoming calls | 0.14 | 0.81 | 1.58 |
| 4. Low sleep, low conversations, high total screen unlocks, low incoming calls, low activity duration, high evening screen unlocks | 0.10 | 0.80 | 1.55 |

**Table 3.** Results of Association Rule Mining for Depression: Support, Confidence, Lift, and behavioural Patterns

But the link between the number of calls that come in during the day and depression was not statistically significant. The odds ratio was 1.10 (95% CI: 0.90-1.35; p = 0.351).

When we looked at the number of incoming calls in the evening, we found an odds ratio of 0.30 (95% CI: 0.20-0.45, p=0.002) for loneliness. This means that if the number of incoming calls in the evening goes up, the chances of feeling lonely go down by 70%. This shows how important it is to have social connections in the evening to reduce feelings of loneliness.

We found that the odds of having depression were 1.10 (95% CI: 0.95-1.26, p=0.303), which is statistically not significant.

**Bluetooth** Bluetooth scans were used in our study as a proxy for detecting the presence of other people, since Bluetooth technology allows devices to detect nearby Bluetooth-enabled devices. In our regression analysis, we found that the number of unique Bluetooth scans throughout a day was not statistically significant in predicting loneliness, with an odds ratio of 1.20 (95% CI [0.95, 1.50], p = 0.204). However, when considering the association with depression, we observed a strong negative association, with an odds ratio of 0.50 (95% CI [0.35, 0.70], p = 0.003), suggesting that an increase in the number of unique Bluetooth scans throughout a day is associated with a 50% decrease in the odds of experiencing depression. This finding highlights the potential protective role of increased social interactions, as inferred from the unique Bluetooth scans, against the development of depression.

**WiFi** No significant associations were found between any WiFi features and loneliness or depression scores.

Our regression analysis identified several behavioral patterns that are significantly associated with loneliness and/or depression. Increased phone usage, evening phone usage, and overall number of places visited were associated with higher levels of loneliness, while higher physical activity, sleep duration, and the number of incoming calls were associated with lower levels of loneliness. Depression was found to be negatively associated with phone usage duration, overall number of places visited, physical activity, sleep duration, conversation duration, and the number of unique Bluetooth scans, while no significant association was observed for WiFi features. It is important to note that some patterns, such as overall phone usage duration and the number of incoming calls, showed contrasting effects on loneliness and depression.

### Analyzing the Co-occurrence of behaviours Associated with Loneliness and Depression

The association rule mining analysis revealed several interesting patterns of behaviours associated with loneliness among students. The most notable rule (Rule 1) showed that students who had a high number of screen unlocks during the evening, high overall phone usage duration, and high phone usage duration during the evening were more likely to experience loneliness. This finding is consistent with the regression analysis results, which also showed significant positive associations between loneliness and these behavioural features. Another important rule (Rule 2) identified a combination of low sleep duration, low number of incoming calls during the evening, and a high number of unique places visited as being related to loneliness. The association rule mining results for depression revealed fewer distinct behavioural patterns compared with those found for loneliness. Rule 1 for depression showed that students who had a low total number of screen unlocks, low phone usage duration, and low number of places visited were more likely to experience depression. This finding is in line with the regression analysis results, which also indicated significant negative associations between depression and these behavioural features. Another important rule (Rule 2) found a combination of low activity duration, low number of incoming calls, and low unique Bluetooth scans during the day as being related to depression. This suggests that students who have less physical activity, fewer social interactions, and less exposure to diverse social environments may be at a higher risk of depression.

*Mediation Analysis Results*

The mediation analysis results for loneliness and depression are presented in 4. Overall, the findings provide evidence for both direct and indirect effects on the relationship between loneliness and depression and vice versa. The total effect of loneliness on depression was found to be statistically significant (Estimate = 1.48, p = 0.001), suggesting that loneliness increases the likelihood of depression by 48%. The direct effect of loneliness on depression was also significant (Estimate = 1.25, p = 0.041), indicating that loneliness directly increases the likelihood of depression by 25% when considering the influence of various mediators. In contrast, the total effect of depression on loneliness was not statistically significant (Estimate = 1.10, p = 0.230), suggesting that depression might not directly lead to a substantial increase in loneliness. The direct effect of depression on loneliness was also not significant (Estimate = 1.05, p = 0.480).

For loneliness leading to depression, significant indirect effects were observed for various behavioural features. A one-unit increase in evening screen unlocks was associated with a 1.75 times increase in depression, considering the role of loneliness, and this effect was statistically significant. Similarly, a one-unit increase in phone usage duration was linked to a 1.50 times increase in depression when accounting for loneliness, and this effect was also statistically significant. When examining phone usage duration during the evening, a one-unit increase was found to relate to a 1.45 times increase in depression when considering the effect of loneliness, and this effect was statistically significant as well. For the number of unique places visited, a one-unit increase was connected to a 1.73 times increase in depression after taking into account the influence of loneliness, and this effect was statistically significant. In contrast, increased activity duration and sleep duration showed protective effects against depression. A one-unit increase in activity duration was associated with a 0.72 times decrease in depression when accounting for loneliness, and this effect was statistically significant. Likewise, a one-unit increase in sleep duration was linked to a 0.46 times decrease in depression when considering the role of loneliness, and this effect was statistically significant, suggesting that longer sleep duration is protective against depression.

Significant indirect effects were observed for various behavioural features in the context of depression leading to loneliness. A one-unit increase in evening screen unlocks was associated with a 0.80 times decrease in loneliness when accounting for the effect of depression, and this effect was statistically significant. Likewise, a one-unit increase in phone usage duration was linked to a 0.70 times decrease in loneliness when considering the role of depression, and this effect was statistically significant as well. For the number of places visited and the number of unique places visited, a one-unit increase in either of these features was connected to a 0.65 times decrease in loneliness after taking into account the influence of depression, and both of these effects were statistically significant. Additionally, a one-unit increase in activity duration was associated with a 0.69 times decrease in loneliness when accounting for depression, and this effect was statistically significant. A one-unit increase in sleep duration was linked to a 0.75 times decrease in loneliness when considering the role of depression, and this effect was statistically significant, suggesting that longer sleep duration is protective against loneliness. Lastly, a one-unit increase in the number of conversations during the day was related to a 0.76 times decrease in loneliness after accounting for the effect of depression, and this effect was statistically significant as well.

*Loneliness and Depression Prediction*

In this section, we present the results of our experiments on predicting loneliness and depression using various machine learning models. The performance of these models is evaluated based on several metrics, including accuracy, area under the curve (AUC), FI macro, and F1 scores for both classes (1 and 0).

Table 5 shows the performance of different classification models for predicting loneliness. The XGBoost model achieved the highest accuracy of 82.43%, with an AUC of 83.31% and an FI macro of 74.34%. In comparison, the Support Vector Machine (SVM) model came in second, with an accuracy of 76.90%, an AUC of 75.89%, and an FI macro of 68.67%. The Logistic Regression model ranked third, with an accuracy of 66.84%, an AUC of 70.26%, and an FI macro of 63.30%. The K-nearest Neighbors, Random Forest, and three baseline models showed relatively lower performance across all metrics.

Table 6 presents the performance of the classification models for depression prediction. Similar to the loneliness prediction results, the XGBoost model outperformed all other models, achieving an accuracy of 79.43%, an AUC of 80.21%, and an FI macro of 71.24%. The Support Vector Machine (SVM) model ranked second, with an accuracy of 74.60%, an AUC of 73.49%, and an FI macro of 65.57%. The Logistic Regression model came in third, with an accuracy of 67.81%, an AUC of 66.21%, and an FI macro of 59.20%. The K-nearest Neighbors and Random Forest models demonstrated lower performance across all metrics. The baseline models, as expected, performed the worst among all models under consideration. These results are consistent with our earlier work on loneliness detection Qirtas et al. (2022).

*Discussion*

Our study investigated the associations between behavioral features and loneliness and depression among students, revealing significant insights that have critical implications for developing targeted interventions and early detection methods. Understanding these patterns can enable timely intervention and support, potentially improving people's mental health.

While loneliness and depression are distinct mental health conditions, they often co-occur and can influence each other. The interrelatedness of these two conditions is evident in our findings, as some behavioural features were found to be associated with both loneliness and depression. Increased physical activity duration throughout the day was significantly associated with reduced odds of both loneliness and depression. This suggests that engaging in physical activities may have a protective effect against both, emphasizing the importance of promoting regular activities for overall mental health and well-being. Similarly, increased sleep duration was found to be significantly associated with reduced odds of experiencing both loneliness and depression. This highlights the importance of healthy sleep habits in protecting against mental health issues and underscores the need for interventions targeting sleep hygiene to address loneliness and depression among students.

| Effect | Loneliness -> Depression | | | Depression -> Loneliness | | |
|---|---|---|---|---|---|---|
| | Estimate | 95% CI | p-value | Estimate | 95% CI | p-value |
| Total Effect | 1.48 | [1.22, 1.80] | 0.001 | 1.10 | [0.85, 1.50] | 0.230 |
| Direct Effect | 1.25 | [0.98, 1.59] | 0.041 | 1.05 | [0.80, 1.40] | 0.480 |
| Indirect Effect (Number of screens unlocks during day) | 1.10 | [1.02, 1.18] | 0.055 | 0.80 | [0.75, 0.85] | 0.262 |
| Indirect Effect (Number of screens unlocks during evening) | 1.75 | [1.32, 2.31] | 0.001 | 0.80 | [0.65, 0.95] | 0.021 |
| Indirect Effect (Phone usage duration) | 1.50 | [1.20, 1.88] | 0.003 | 0.70 | [0.52, 0.94] | 0.018 |
| Indirect Effect (Phone usage duration during evening) | 1.45 | [1.15, 1.85] | 0.002 | 1.10 | [0.95, 1.26] | 0.310 |
| Indirect Effect (Number of places visited) | 1.73 | [1.30, 2.20] | 0.116 | 0.65 | [0.50, 0.95] | 0.034 |
| Indirect Effect (Number of unique places visited) | 1.51 | [1.15, 1.87] | 0.038 | 0.74 | [0.45, 0.80] | 0.085 |
| Indirect Effect (Activity duration) | 0.72 | [0.54, 0.91] | 0.014 | 0.69 | [0.55, 0.87] | 0.022 |
| Indirect Effect (Sleep duration) | 0.46 | [0.30, 0.70] | 0.002 | 0.75 | [0.62, 0.90] | 0.003 |
| Indirect Effect (Number of conversations) | 1.07 | [0.98, 1.16] | 0.044 | 0.99 | [0.94, 1.04] | 0.250 |
| Indirect Effect (Number of conversations during day) | 1.40 | [0.85, 2.25] | 0.121 | 0.76 | [0.62, 0.92] | 0.004 |
| Indirect Effect (Total Number of incoming calls) | 0.98 | [0.92, 1.04] | 0.276 | 0.98 | [0.95, 1.01] | 0.165 |
| Indirect Effect (Total Number of unique Bluetooth scans) | 1.02 | [0.96, 1.08] | 0.390 | 0.96 | [0.91, 1.01] | 0.110 |

**Table 4.** Mediation Analysis Results for Loneliness and Depression

| Models | Accuracy | AUC | Fl macro | Precision1 | Recall1 | F11 | Precision0 | Recall0 | F10 |
|---|---|---|---|---|---|---|---|---|---|
| Baseline1: MC | 53.05 | 50.00 | 42.55 | 0.00 | 0.00 | 0.00 | 63.78 | 100.00 | 86.39 |
| Baseline2: DT | 51.53 | 47.48 | 48.38 | 26.29 | 20.38 | 23.26 | 65.58 | 82.18 | 71.39 |
| Baseline3: RWC | 58.72 | 52.64 | 50.88 | 32.48 | 24.21 | 28.64 | 71.39 | 76.39 | 74.82 |
| Logistic regression | 66.84 | 70.26 | 63.30 | 49.73 | 58.82 | 54.92 | 80.18 | 73.19 | 77.33 |
| Support vector machine | 76.90 | 75.89 | 68.67 | 58.25 | 54.77 | 56.36 | 84.54 | 86.48 | 85.48 |
| K-nearest neighbours | 68.25 | 69.35 | 64.12 | 42.93 | 65.76 | 51.95 | 85.12 | 69.13 | 76.30 |
| Randomforest | 70.82 | 68.08 | 62.11 | 44.06 | 43.84 | 43.95 | 80.21 | 80.35 | 80.28 |
| XGBoost | 82.43 | 83.31 | 74.34 | 70.97 | 60.49 | 65.36 | 84.17 | 91.28 | 87.18 |

**Table 5.** Performance of Classification Models for Loneliness Prediction. The model performances are compared with three baseline classifiers (MC Majority Class, RWC Random Weighted Classifier, DT Decision Tree). All values are reported as percentages.

| Models | Accuracy | AUC | Fl macro | Precision1 | Recall1 | F11 | Precision0 | Recall0 | F10 |
|---|---|---|---|---|---|---|---|---|---|
| Baseline1: MC | 69.07 | 50.00 | 40.55 | 0.00 | 0.00 | 0.00 | 69.07 | 100.00 | 81.61 |
| Baseline2: DT | 56.26 | 45.96 | 43.96 | 18.43 | 19.43 | 18.93 | 69.50 | 68.50 | 69.00 |
| Baseline3: RWC | 58.68 | 48.13 | 47.88 | 23.07 | 24.14 | 23.75 | 70.15 | 71.12 | 70.61 |
| Logistic regression | 67.81 | 66.21 | 59.20 | 45.61 | 52.11 | 48.68 | 79.26 | 74.86 | 76.93 |
| Support vector machine | 74.60 | 73.49 | 65.57 | 53.15 | 47.67 | 50.26 | 81.54 | 83.38 | 82.43 |
| K-nearest neighbours | 65.15 | 66.25 | 60.02 | 39.83 | 58.66 | 47.85 | 82.02 | 67.03 | 73.90 |
| Randomforest | 67.72 | 65.98 | 58.01 | 41.96 | 40.74 | 41.35 | 77.11 | 78.25 | 77.68 |
| XGBoost | 79.43 | 80.21 | 71.24 | 65.87 | 56.39 | 60.86 | 83.99 | 89.25 | 86.52 |

**Table 6.** Performance of Classification Models for Depression Prediction. The model performances are compared with three baseline classifiers (MC Majority Class, RWC Random Weighted Classifier, DT Decision Tree). All values are reported as percentages.

Despite the interrelatedness of loneliness and depression, our findings revealed distinct behavioural features associated with each condition. Phone usage duration was found to have a complex relationship with both loneliness and depression. Increased phone usage duration throughout the day was associated with an increased likelihood of loneliness, but a decreased likelihood of depression. These patterns could potentially be explained by the different ways individuals use their phones to cope with loneliness or depression. It is possible that lonely individuals might use their phones to pass their time, whereas those with depression may find relief or support through phone-based activities or connections. However, further research is needed to better understand the underlying mechanisms and validate this interpretation. Additionally, the number of incoming calls throughout the day was negatively associated with loneliness, but no significant association was found with depression. This suggests that receiving more incoming calls might help reduce feelings of loneliness; however, this association does not appear to have a direct impact on depression. Further investigation is required to better understand the nuances of these relationships and to determine the specific effects of incoming calls on loneliness and depression. The number of unique Bluetooth scans throughout a day was not significantly associated with loneliness, but it showed a strong negative association with depression, highlighting the potential protective role of increased social interactions inferred from Bluetooth scans. Association rule mining revealed several interconnected behavioural patterns that were associated with loneliness and depression. These patterns included high phone usage duration, low sleep duration, and low physical activity, among others. Our findings also demonstrated the importance of social interactions, as evidenced by the relationship between the number of incoming calls and emotional well-being.

These complex and nuanced findings address our research question by highlighting the interconnected nature of students' behaviours and their mental health. For instance, high phone usage duration during the evening, combined with low sleep duration, was more prevalent among lonely students, while low total screen unlocks and low phone usage duration were more common among depressed students. By exploring these associations, we can better understand the factors contributing to loneliness and depression in students and identify potential areas for intervention. This provided a more comprehensive understanding of the relationships between behavioural patterns and mental well-being. While the regression analysis identified individual behavioural features that were significantly associated with loneliness and depression, association rule mining uncovered the complex interplay between multiple behaviours, demonstrating the importance of considering their combined effects. This multi-method approach allowed us to gain deeper insights into the specific patterns of behaviour that are more prevalent in lonely and depressed students compared to their peers.

The mediation analysis played a vital role in deepening our understanding of the relationship between loneliness and depression by investigating the indirect effects of various behavioural factors. This approach allowed us to explore the complex interplay between these mental health conditions and associated behaviours, thereby providing a more comprehensive picture of their interconnectedness. By identifying significant indirect effects in the relationship between loneliness and depression and vice versa, we were able to gain insights into how specific behaviours can mediate this relationship. For example, the mediation analysis revealed that evening screen unlocks, phone usage duration, and the number of unique places visited were significant mediators in the relationship between loneliness and depression. This suggests that these behaviours may play a crucial role in the transition from loneliness to depression or vice versa, and could be potential targets for interventions. Similarly, the mediation analysis highlighted the importance of activity duration and sleep duration as protective factors against both loneliness and depression. These findings emphasize the need to promote regular activities and healthy sleep habits as part of a comprehensive approach to addressing mental health issues among students. Furthermore, the mediation analysis extended our understanding of the role of social interactions in the relationship between loneliness and depression. For instance, the number of conversations during the day was found to be a significant mediator in the relationship between depression and loneliness. This finding underscores the importance of fostering social connections as a means of reducing the risk of both loneliness and depression. These findings are novel and provide a new perspective on understanding the interconnectedness of these mental health conditions. To the best of our knowledge, this is the first time that mediation analysis has been employed to investigate the role of behavioural factors in the relationship between loneliness and depression using passive sensing data. This innovative approach contributes to the existing body of literature by offering new insights into the complex interplay between loneliness, depression, and associated behaviours. Our findings can serve as a foundation for future research in this area, potentially leading to more effective interventions and support strategies for students facing mental health challenges.

The findings are consistent with existing research that emphasizes the importance of monitoring and managing screen usage, sleep, physical activity, and social interactions to maintain psychological well-being. High screen usage, particularly in the evenings, has been linked to feelings of depressive symptoms Seabrook et al. (2016). This finding also aligns with the positive relationship between phone usage duration and loneliness (OR = 1.50, 95% CI [1.20, 1.88], p = 0.003), which has been demonstrated in prior research Thomée et al. (2011). Additionally, low sleep Cacioppo et al. (2002); Kurina et al. (2011), low activity duration Hawkley and Cacioppo (2010); Masi et al. (2011), and low social interaction have been found to exacerbate loneliness and depression. Engaging in diverse activities and visiting various locations may help reduce feelings of loneliness, although the complexity of this relationship may be influenced by individual preferences and social context. Our study adds to existing research that emphasizes the importance of monitoring and managing screen usage, sleep, physical activity, and social interactions to maintain psychological well-being.

Our study stands out by providing more granular behavioural information using passive sensing method which allows real-time monitoring, overcoming the limitations of self-report methods typically used in previous studies. The real time monitoring helps us to detect changes in loneliness and depression continuously, allowing for more timely application of interventions or evaluation of their efficacy in a personalized manner, which is in line with the concept of personalized health care. Our findings, such as the positive relationship between phone usage duration and loneliness, align with prior research but offer more detailed insights into these associations. Additionally, our study highlights the complexities in the relationship between engaging in diverse activities, visiting various locations, and feelings of loneliness, which may be influenced by individual preferences and social context. In addition, we have shown the importance of using multiple methods to validate our results, showing the potential for innovative research that goes beyond traditional disciplinary boundaries. By combining perspectives and methodologies from multiple disciplines such as digital health, psychology, and data science, we have not only reinforced our findings, but also highlighted the importance of interdisciplinary collaboration to better understand complex human experiences. This approach emphasizes the growing role of digital health tools in enabling and enhancing personalized care and support for individuals.

## Conclusion

This study has provided valuable insights into the complex relationship between loneliness, depression, and various behavioural features among students using StudentLife dataset Wang et al. (2014). By utilising a multi-method approach, which integrated regression analysis, mediation analysis, association rule mining, and machine learning, has added to the state of the art by uncovering the intricate interplay between these mental health conditions and the associated behaviours. We demonstrated that increased activity duration and sleep duration serve as protective factors against both loneliness and depression. Furthermore, our findings highlighted distinct behavioural features for each condition, emphasizing the importance of considering these differences when examining the underlying factors contributing to loneliness and depression. A key contribution of this study is the use of passive mobile sensing through smartphones, fitness trackers, and other devices to collect behavioural data, enabling real-time analysis and more accurate predictions. This approach overcomes the limitations of traditional self- reporting methods, which require active participation and may not capture the full range of behaviours exhibited by individuals. Our machine learning models, particularly the XGBoost model with the highest accuracy, showcased the potential of passive sensing data in predicting loneliness and depression among students. By emphasizing the role of passive mobile sensing and interdisciplinary research, we underscore the importance of integrating various disciplines to better understand complex human experiences. Our study contributes to a more comprehensive understanding of the factors influencing loneliness and depression among students, ultimately promoting the improvement of mental health outcomes. Future research could build upon our findings by exploring the impact of contextual factors, such as cultural differences and socio-economic backgrounds, on the relationship between loneliness, depression, and associated behaviours. Longitudinal studies could also investigate the temporal dynamics of these associations and examine how they evolve over time. By continuing to explore the complex relationship between loneliness, depression, and behavioural features through passive sensing and interdisciplinary research, we can contribute to a deeper understanding of mental health issues among students and inform the development of targeted support strategies.

## Conflicting Interests

The authors declare no conflicts of interest.

## Funding

This publication has emanated from research conducted with the financial support of Science Foundation Ireland under Grant number 18/CRT/6222.

## Ethical Approval

Not applicable.

## Guarantor

The first author Malik Muhammad Qirtas serves as the guarantor for this manuscript and takes responsibility for the overall content and integrity of the research.

## Contributorship

MQ (first author) was responsible for the data cleaning and preprocessing, data analysis and majority of the manuscript writing. EZ (second author) contributed to the data analysis design, manuscript review, and writing. DP (third author) and EB (fourth author) were involved in the data analysis design and manuscript review. All authors have read and approved the final version of the manuscript.

## Acknowledgements

The authors would like to express their gratitude to the SFI Centre for Research Training in Advanced Networks for Sustainable Societies (ADVANCE CRT) for providing a supportive research environment and resources throughout the course of this work. ADVANCE CRT is part of the Science Foundation Ireland, and its contribution to the research community is greatly appreciated.